\documentclass[letterpaper,twocolumn,prl,aps,superscriptaddress,amsmath,amssymb,floatfix]{revtex4-1}
\usepackage{mathptmx}
\usepackage[latin9]{inputenc}
\usepackage{color}
\usepackage{amsmath}
\usepackage{amssymb}
\usepackage{graphicx}
\usepackage{esint}
\usepackage[unicode=true,
 bookmarks=true,bookmarksnumbered=false,bookmarksopen=false,
 breaklinks=false,pdfborder={0 0 1},backref=false,colorlinks=true]
 {hyperref}
\hypersetup{
 linkcolor=magenta,urlcolor=blue,citecolor=blue,pdfstartview={FitH},hyperfootnotes=false}

\makeatletter

\usepackage{textcomp}
\usepackage{epstopdf}
\usepackage{xcolor}

\pdfpageheight\paperheight
\pdfpagewidth\paperwidth

\@ifundefined{textcolor}{}{%
 \definecolor{BLACK}{gray}{0}
 \definecolor{WHITE}{gray}{1}
 \definecolor{RED}{rgb}{1,0,0}
 \definecolor{GREEN}{rgb}{0,1,0}
 \definecolor{BLUE}{rgb}{0,0,1}
 \definecolor{CYAN}{cmyk}{1,0,0,0}
 \definecolor{MAGENTA}{cmyk}{0,1,0,0}
 \definecolor{YELLOW}{cmyk}{0,0,1,0}
}
\usepackage{marvosym}
\usepackage{xcolor}\usepackage{soul}
\definecolor{blue}{rgb}{0,0,1}
\definecolor{red}{rgb}{1,0,0}
\definecolor{green}{rgb}{0,1,0}

\usepackage{soul}

\makeatother

\begin{document}

\title{On-chip Parametric Amplification in a Double Quantum Dots Circuit}
%Microwave Readout with an On-chip Semiconductor Parametric Amplifier

\author{Yong-Qiang Xu}
\thanks{These authors contributed equally to this work.}
\affiliation{Laboratory of Quantum Information, University of Science and Technology of China, Hefei 230026, China}
\affiliation{CAS Center for Excellence in Quantum Information and Quantum Physics, University of Science and Technology of China, Hefei 230026, China}

\author{Rui Wu}
\thanks{These authors contributed equally to this work.}
\affiliation{Laboratory of Quantum Information, University of Science and Technology of China, Hefei 230026, China}
\affiliation{CAS Center for Excellence in Quantum Information and Quantum Physics, University of Science and Technology of China, Hefei 230026, China}

\author{Si-Si Gu}
\affiliation{Laboratory of Quantum Information, University of Science and Technology of China, Hefei 230026, China}
\affiliation{CAS Center for Excellence in Quantum Information and Quantum Physics, University of Science and Technology of China, Hefei 230026, China}

\author{Shun-Li Jiang}
\affiliation{Laboratory of Quantum Information, University of Science and Technology of China, Hefei 230026, China}
\affiliation{CAS Center for Excellence in Quantum Information and Quantum Physics, University of Science and Technology of China, Hefei 230026, China}

\author{Shu-Kun Ye}
\affiliation{Laboratory of Quantum Information, University of Science and Technology of China, Hefei 230026, China}
\affiliation{CAS Center for Excellence in Quantum Information and Quantum Physics, University of Science and Technology of China, Hefei 230026, China}

\author{Bao-Chuan Wang}
\affiliation{Laboratory of Quantum Information, University of Science and Technology of China, Hefei 230026, China}
\affiliation{CAS Center for Excellence in Quantum Information and Quantum Physics, University of Science and Technology of China, Hefei 230026, China}

\author{Hai-Ou Li}
\affiliation{Laboratory of Quantum Information, University of Science and Technology of China, Hefei 230026, China}
\affiliation{CAS Center for Excellence in Quantum Information and Quantum Physics, University of Science and Technology of China, Hefei 230026, China}
\affiliation{Hefei National Laboratory, University of Science and Technology of China, Hefei 230088, China}

\author{Guang-Can Guo}
\affiliation{Laboratory of Quantum Information, University of Science and Technology of China, Hefei 230026, China}
\affiliation{CAS Center for Excellence in Quantum Information and Quantum Physics, University of Science and Technology of China, Hefei 230026, China}
\affiliation{Hefei National Laboratory, University of Science and Technology of China, Hefei 230088, China}

\author{Chang-Ling Zou}
\email{clzou321@ustc.edu.cn}
\affiliation{Laboratory of Quantum Information, University of Science and Technology of China, Hefei 230026, China}
\affiliation{CAS Center for Excellence in Quantum Information and Quantum Physics, University of Science and Technology of China, Hefei 230026, China}
\affiliation{Hefei National Laboratory, University of Science and Technology of China, Hefei 230088, China}

\author{Gang Cao}
\email{gcao@ustc.edu.cn}
\affiliation{Laboratory of Quantum Information, University of Science and Technology of China, Hefei 230026, China}
\affiliation{CAS Center for Excellence in Quantum Information and Quantum Physics, University of Science and Technology of China, Hefei 230026, China}
\affiliation{Hefei National Laboratory, University of Science and Technology of China, Hefei 230088, China}
\thanks {*Corresponding author}

\author{Guo-Ping Guo}
\email{gpguo@ustc.edu.cn}
\affiliation{Laboratory of Quantum Information, University of Science and Technology of China, Hefei 230026, China}
\affiliation{CAS Center for Excellence in Quantum Information and Quantum Physics, University of Science and Technology of China, Hefei 230026, China}
\affiliation{Hefei National Laboratory, University of Science and Technology of China, Hefei 230088, China}
\affiliation{Origin Quantum Computing Company Limited, Hefei 230088 , China }

\date{\today}

\begin{abstract}
In microwave-based quantum circuits, including double quantum dots (DQDs), superconducting qubits and spin qubits, parametric amplifiers are indispensable in achieving high-fidelity qubit readouts. Despite its importance, the application of parametric amplifiers is hampered by several challenges, such as high insertion losses, constrained tunability, and a pronounced vulnerability to magnetic fields. Here, we demonstrate an on-site single-atom parametric amplifier (SAPA) within a reconfigurable quantum circuit, which consists of a superconducting microwave cavity and two GaAs gate-defined DQDs. Leveraging the inherent nonlinearity of the DQD, a parametric gain exceeding 11\,dB is achieved. This gain contributes to enhance the qubit readout, as evidenced by exceeding two times improvement in the signal-to-noise ratio (SNR) when employing the DQD-based amplifier for reading out another DQD. Our work not only presents a versatile experimental platform with enhanced readout capabilities in quantum computing, but also introduces alternative choices of parametric amplifiers for a variety of microwave-based quantum circuits.

\end{abstract}

\maketitle

\section{Introduction}\label{section1}

Superconducting solid-state microwave cavities have emerged as a crucial platform for scalable quantum information processing, thanks to their ability to couple with various types of qubits, such as superconducting qubits~\cite{Wallraff_Strong_2004,Majer_Coupling_2007}, double quantum dots (DQDs)~\cite{Mi_Strong_2017,Stockklauser_Strong_2017,Wang_Correlated_2021,Kang_Coupling_2024}, and spins~\cite{Kubo_Strong_2010,Tabuchi_Coherent_2015,Viennot_Coherent_2015,Zhang_Cavity_2015,Landig_Coherent_2018,Mi_A_2018,Samkharadze_Strong_2018,Yu_Strong_2023,Ungerer_Strong_2024}. These microwave cavities play a vital role in mediating long-range coupling between qubits~\cite{Majer_Coupling_2007,Wang_Correlated_2021,Woerkom_Microwave_2018,Borjans_Resonant_2020,Bottcher_Parametric_2022,Dijkema_Two_2024} and the distribution of quantum information~\cite{Zhong_Deterministic_2021,Zhou2024a}. Among the diverse qubit platforms, the DQD-microwave cavity system stands out as a particularly promising candidate for versatile applications~\cite{Xu_Coupling_2023}, owing to its unique advantages, including large coupling strength~\cite{Scarlino_In_2022,Yu_Strong_2023,Janik_Strong_2025}, excellent tunability via electric fields, and compatibility with other qubit architectures~\cite{Scarlino_Coherent_2019}. Integrating DQDs with microwave cavities offers unparalleled flexibility and scalability~\cite{Vandersypen_Interfacing_2017,Holman_3D_2021,Li_Complete_2025}, enabling the realization of strong light-matter interactions~\cite{Bonsen_Probing_2023}, coherent quantum phenomena~\cite{Gu_Probing_2023,Dijkema_Two_2024}, and non-trivial functionalities like microwave photon sources~\cite{Liu_Semiconductor_2015,Gu_Gain_2023}.

Meanwhile, to achieve high-fidelity readout of qubit states through microwave photons, it is essential to use low-temperature, low-noise parametric amplifiers. Conventional readout techniques often suffer from low signal-to-noise ratios (SNRs) and high insertion losses. Previous efforts have focused on cascading additional amplifiers, such as Josephson parametric amplifiers (JPAs)~\cite{Castellanos_Widely_2007} and traveling-wave parametric amplifiers (TWPAs)~\cite{Macklin_A_2015,Zhang_Traveling_2023}, to enhance the readout signal. However, these separate devices introduce extra insertion losses and are often sensitive to magnetic fields and frequency mismatches, limiting their practicality and performance. To overcome these limitations, on-site amplifiers that can be seamlessly integrated with quantum devices offer an attractive alternative, particularly for improving readout fidelity~\cite{Wen_Reflective_2018,Cochrane_Parametric_2022,Vine_In_2023}. 

Here, inspired by the concept of the single two-level atom amplifier~\cite{Mollow_Stimulated_1972,Wu_Observation_1977}, we demonstrate an on-site fine-tunable single-atom parametric amplifier (SAPA) based on DQD.  By integrating two DQDs into a high-impedance microwave cavity, we develop a reconfigurable hybrid quantum circuit. Leveraging the strong nonlinearity of a DQD, we achieve a remarkable parametric gain exceeding 11\,dB and enhance the readout of a target DQD with approximately two-fold SNR improvement. The dynamic reconfigurability and adaptability of the system is demonstrated by swapping the roles of the two DQDs between acting as the amplifier and the measured qubit. Notably, the SAPA holds promise for operation under high magnetic fields and breaking the limit to work at higher temperatures ~\cite{Petit_Universal_2020,Yang_Operation_2020}, which is crucial for semiconductor spin qubit-based quantum computing. Our results represent a significant step in the development of scalable and high-performance quantum technologies, paving the way for a new generation of quantum devices with enhanced readout capabilities and novel functionalities.

\section{Device architecture}\label{section2}

\begin{figure}[!t]
	\includegraphics[width = 8.6cm]{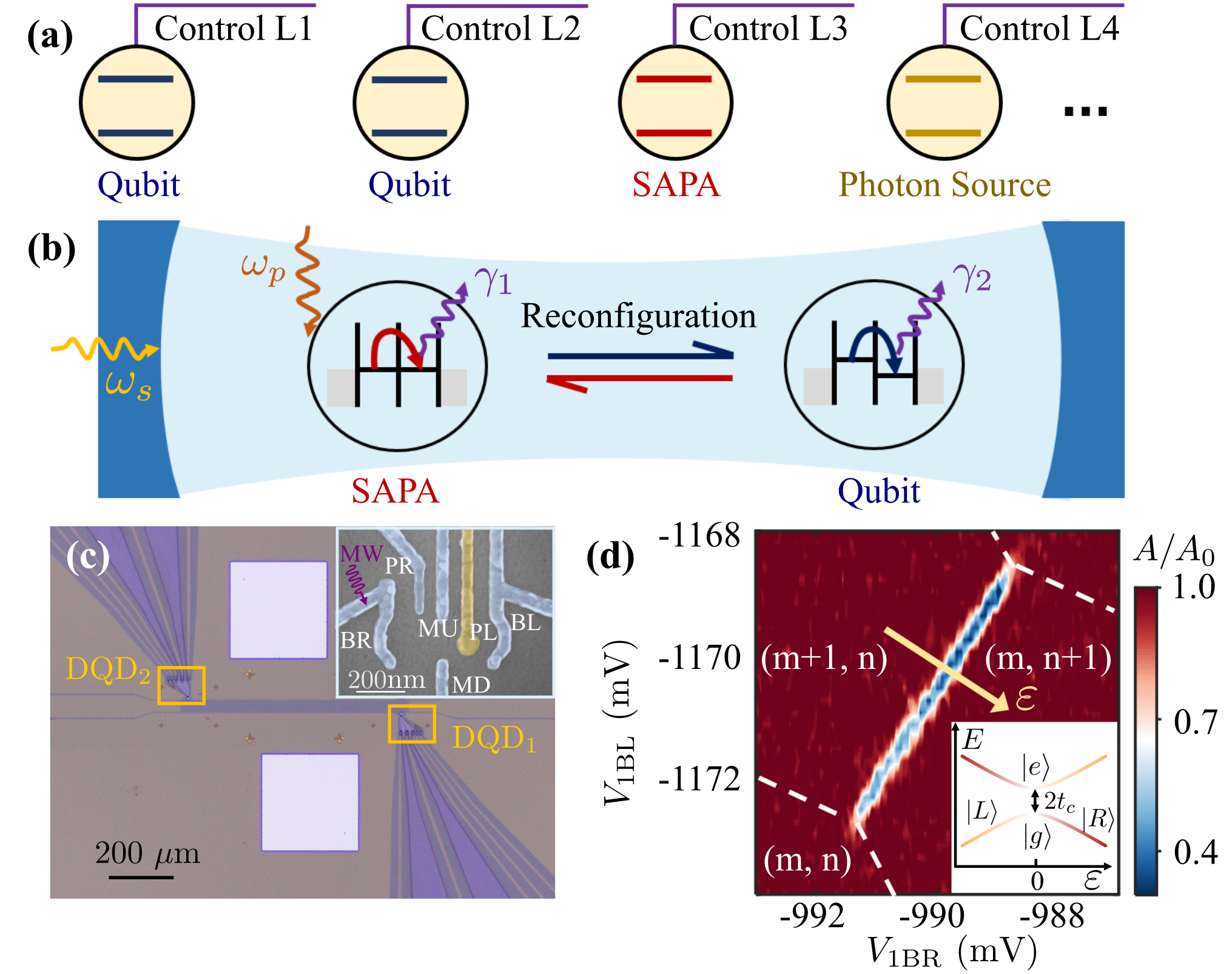}
	\caption{(a) Illustration of the reconfigurable quantum system, where functions of DQDs are adaptable through control lines ``Control Lx". (b) Sketch of parametric amplification and on-chip integrated readout in the hybrid system composed of DQDs and a cavity.  (c) Optical micrograph of the experimental device, which comprises two DQDs coupled to a cavity. Inset: false-color scanning electron micrograph of the DQD. The plunger gate PL (orange) is connected to the cavity. (d) Normalized cavity transmission amplitude $A/A_0$ of DQD$_1$ as functions of gate voltages $V_{\mathrm{1BR}}$ and $V_{\mathrm{1BL}}$, where $A_0$ is the averaged cavity transmission amplitude with DQD$_1$ in the Coulomb blockade regime. The orange arrow indicates the detuning $\varepsilon$, which is adjusted by $V_{\mathrm{1BR}}$ with the lever arm 0.072 in this experiment. Inset: Energy diagram of the charge qubit defined in DQD.}
 \label{Fig1}
	
\end{figure}

The reconfigurable hybrid quantum system illustrated in Fig.~\ref{Fig1}(a) showcases the versatility of a microwave cavity coupled to multiple DQDs. By harnessing the unique properties of DQDs, such as their strong coupling to cavity photons and excellent electrical tunability, DQDs can be reconfigured as qubits, SAPAs, as well as microwave photon sources ~\cite{Liu_Semiconductor_2015,Gu_Gain_2023,Li_Frequency_2023}. This platform offers a rich playground for exploring coherent quantum dynamics, implementing quantum information processing, and developing novel functional devices. To demonstrate the potential of this platform, we implement a proof-of-concept experiment based on the scheme depicted in Fig.~\ref{Fig1}(b). In this setup, two DQDs are coupled to a common microwave cavity, with one DQD serving as a qubit ~\cite{Petersson_Quantum_2010,Cao_Ultrafast_2013}, while the other DQD acts as a SAPA to enhance readout. This arrangement allows for the realization of an on-site integrated readout system that leverages the nonlinearity of DQD to achieve high-fidelity, low-noise amplification.

The hybrid device, shown in Fig.~\ref{Fig1}(c), comprises a half-wavelength transmission microwave cavity and two spatially separated GaAs/AlGaAs DQDs with a distance of $670~\rm\mu m$. The cavity is fabricated from 11-nm-thick NbTiN film, with a high impedance up to $2~{\mathrm k\Omega}$, a center frequency of $\omega_r/2\pi=5.198~\rm GHz$ and a photon decay rate of $\kappa/{2\pi}=14~\rm MHz$. Each DQD is galvanically connected to the cavity via the plunger gate PL. An excess valence electron in the DQD defines the charge states $\left | {L} \right \rangle=\left | {m+1,n} \right \rangle$ and $\left | {R} \right \rangle=\left | {m,n+1} \right \rangle$, where $m~(n)$ represents the occupation number of electrons in the left (right) dot. As illustrated in the inset of Fig.~\ref{Fig1}(d), an anti-crossing occurs between the levels near the energy detuning $\varepsilon=0$, where the gap corresponds to the interdot tunnel coupling $2t_c$ controlled by gate MU and MD. In the basis of $\{ \left | {R} \right \rangle,\,\left | {L} \right \rangle \}$, the Hamiltonian for the DQD is given by $H_{q}=\frac{1}{2}\varepsilon\sigma_{z}+t_{c}\sigma_{x}$, where $\sigma_{z}$ and $\sigma_{x}$ are the Pauli operators. The eigenstates $\left | {g} \right \rangle$ and $\left | {e} \right \rangle$ are denoted as ground and excited states of the qubit, respectively, showing a tunable transition frequency $\omega_q=\sqrt{\varepsilon^2+{4t_c}^2}/\hbar$ ~\cite{Petersson_Quantum_2010, Cao_Ultrafast_2013}. We measure the charge stability diagram of DQD$_1$ by monitoring the cavity transmission amplitude, as shown in Fig.~\ref{Fig1}(d). The observed characteristic honeycomb pattern indicates the transitions of a charge qubit. 

\section{Artificial atom-based parametric amplifier}\label{section3}

The hybrid system can be described as
\begin{equation}
    H=\hbar\omega_ca^{\dagger}a+\sum_{j=1,2}\left[\frac{1}{2}\varepsilon^{(j)}\sigma_{z}^{(j)}+t_{c}^{(j)}\sigma_{x}^{(j)}+\hbar g_c^{(j)}\sigma_{z}^{(j)}(a^{\dagger}+a)\right],\notag
\end{equation}
where  $a(a^{\dagger})$ denotes the annihilation (creation) operator of the cavity mode, and $g_c^{(j)}$ is the coupling strength between the cavity and DQD$_{j}$ with superscripts $j$ denoting $j$-th DQD. Due to the high characteristic impedance $Z_r$ of the cavity, the coupling strength $g_c^{(j)} \propto \sqrt{Z_r}$ is significantly enhanced~\cite{Samkharadze_High_2016,Stockklauser_Strong_2017}. The two DQDs have their energy decay rates $\gamma_{1,2}$, and the system is characterized by external microwave pump and probe fields with frequencies $\omega_p$ and $\omega_s$.

\begin{figure}[!t]
\includegraphics[width = 8.6cm]{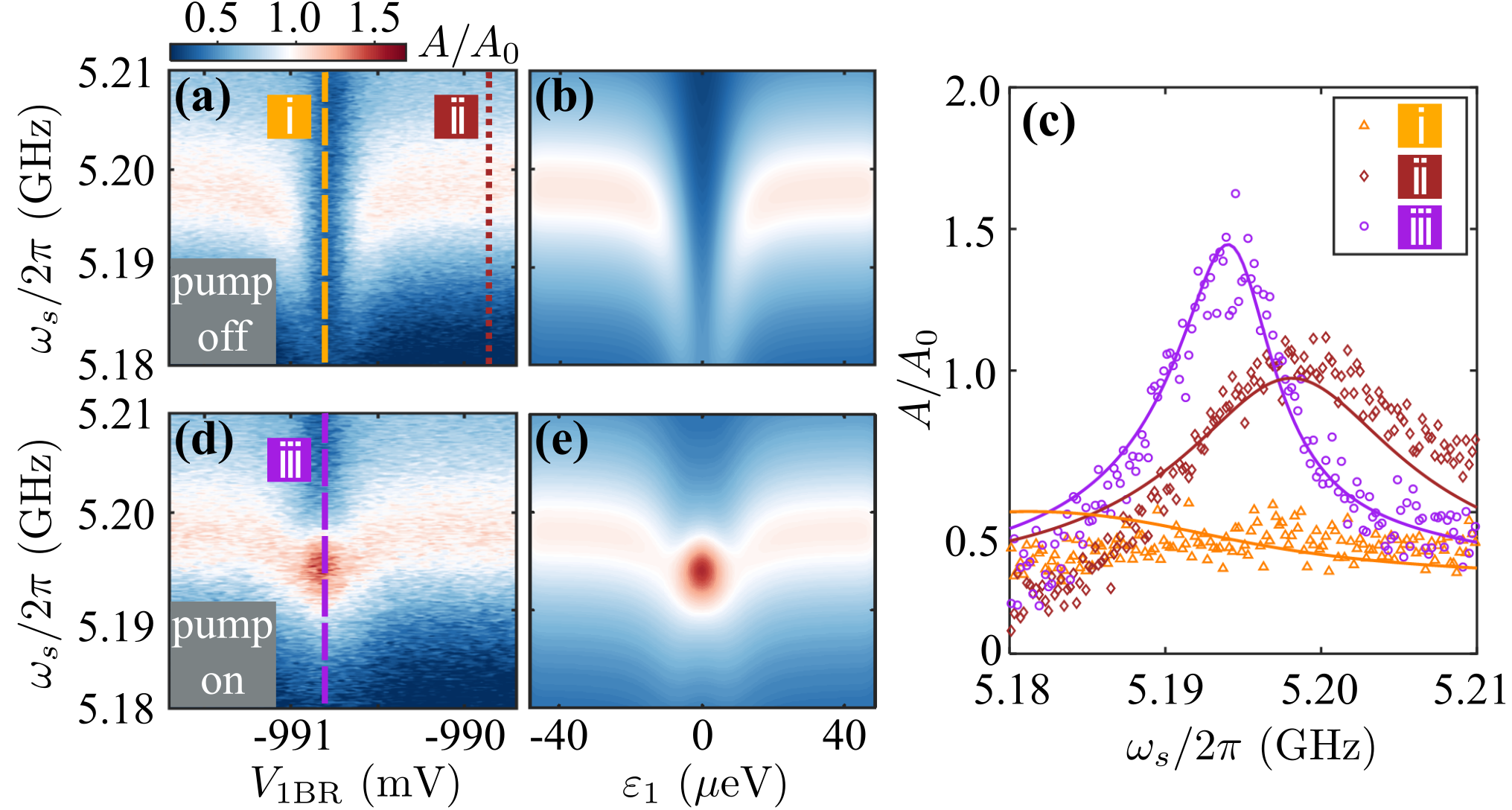} 
\caption{(a) Vacuum Rabi splitting pattern of DQD$_1$ without applying a pump tone. (b) Corresponding simulation result for (a). (c) Comparison of the maximum transmission spectrum with a pump tone applied (purple circles), at the same detuning without a pump tone (orange triangles), and for a large detuning without a pump tone (brown diamonds). The lines are theoretical calculations using independently determined parameters, rather than fitting of the experimental data (symbols). (d) Cavity transmission with the same parameters as in (a) but with a pump tone applied, where the pump-signal detuning $\Delta \omega=2\pi\times5\,\mathrm{kHz}$. (e) Corresponding simulation result for (d).}
 \label{Fig2}
\end{figure}

To investigate the parametric amplification effect, we start from a simplified system with DQD$_2$ isolated by setting a transition frequency that is far off-detuning, while DQD$_1$ is set to $2t_{c}^{(1)}/2\pi\hbar = 5.32~\rm GHz$. In Fig.~\ref{Fig2}(a), we measure the vacuum Rabi splitting spectrum of the strongly coupled DQD and the cavity by sweeping the probe frequency $\omega_s$ and recording the normalized transmission amplitude $A/A_0$ as a function of detuning $\varepsilon$. As denoted by the brown circles in Fig.~\ref{Fig2}(c), the cavity features a Lorentzian-type transmission spectrum when DQD$_1$ is largely off-resonant, as expected from a typical empty cavity. In contrast, as shown by the orange triangles, the amplitude is significantly suppressed when it is near resonance $\varepsilon=0$ ~\cite{Viennot_Coherent_2015, Blais_Circuit_2021}. The observed noise floor is attributed to a combination of intrinsic device charge noise and measurement chain noise. We fit the experimental results by solving the equations of motion for operator expectation and input-output theory~\cite{Collett_Squeezing_1984,Lalumiere_Input_2013}. In Fig.~\ref{Fig2}(b), the results are in good agreement with the experimental data, giving a coupling strength $g_{c}^{(1)}/{2\pi}\approx60~\rm MHz$ and a decoherence rate $\gamma_{1}/{2\pi}\approx100~\rm MHz$. The strong coupling is corroborated by the high cooperativity $C=4[g_c^{(1)}]^2 /\gamma_1\kappa\approx10$~\cite{Clerk_Hybrid_2020} and the criterion $g_{c}^{(1)}>\gamma_1/2,\kappa/2$, indicating the coherent dynamics dominates dissipative processes, enabling efficient energy exchange and information transfer between the DQD and the cavity.

The strong coupling and high cooperativity of the DQD-cavity system form the basis for realizing a SAPA. The operating principle of the amplifier relies on the nonlinearity of DQD, which induces a four-wave mixing process in the presence of a strong pump tone~\cite{Savelev_Two_2012,Pirkkalainen_Cavity_2015,Wen_Reflective_2018}. When the pump frequency $\omega_p$ is properly chosen, the nonlinear interaction between the pump, signal, and idler photons leads to the signal amplification at frequency $\omega_s$. This process is analogous to parametric amplification in conventional nonlinear media, but here it is achieved with a single artificial atom (DQD$_1$) coupled to a microwave cavity. To demonstrate the parametric amplification effect, we activate the pump tone $\omega_p$ with a power $P_{\rm d,1}=-68.3~\rm dBm$ applied in room temperature and measure the cavity transmission shown in Fig.~\ref{Fig2}(d). Here, the spectrum is measured by scanning $\omega_s$ while fixing the pump-signal frequency difference $\Delta\omega=\omega_p-\omega_s=2\pi\times5\,\mathrm{kHz}$. The result reveals a significant amplitude enhancement along the line $\varepsilon=0$ [purple circle and dashed line in Fig.~\ref{Fig2}(c) and (d), respectively]. The simulation result, shown in Fig.~\ref{Fig2}(e), is in excellent agreement with the experimental data. The maximum amplitude is observed when the signal frequency is $\omega_s/2\pi=5.194~\rm GHz$, corresponding to a parametric gain $G_{p}=20*{\rm log}_{10}(A_{\rm on,max}/A_{\rm off})$ of 11.28\,dB. Here, $A_{\rm on,max}$ represents the maximum amplitude when the pump is turned on, and $A_{\rm off}$ is the amplitude at the same position without the pump. The observed gain arises from the strong coupling between the qubit and the cavity, and is therefore enhanced only when the pump and probe frequencies are near resonance with the DQD-modified cavity mode. This selective enhancement is a key signature of the DQD-based SAPA and distinguishes it from conventional parametric amplifiers based on bulk nonlinear materials. The simulations also confirm that the observed amplification arises from the nonlinearity of the DQD, rather than any intrinsic nonlinearity of the cavity. In addition, the Heisenberg equation of motion in Supporting Information can directly illustrate the role of the nonlinearity of quantum dots, and some other evidence is summarized in Supporting Information.

Interestingly, the gain not only enhances the amplitude of the modified cavity mode but also reduces its linewidth. This linewidth narrowing effect is a direct consequence of the increased cavity quality factor due to the presence of the strongly coupled DQD. The simultaneous enhancement of the peak amplitude and the reduction of the linewidth would lead to a significant improvement in the SNR, which is crucial for practical applications in qubit readout. Moreover, we define the effective gain $G_{e}=20*{\rm log}_{10}(A_{\rm on,max}/A_{\rm 0})$, which compares the maximum amplitude with amplification to the bare cavity amplitude without amplification, reflecting the amplification ability to the cavity signal rather than the input signal. This figure of merit is particularly relevant for applications in circuit QED, where the cavity is used as a measurement tool, and the probe frequency is typically set to the cavity resonance. In our experiment, we achieve an effective gain of 4.22\,dB, indicating the potential of the on-site SAPA for enhancing the readout fidelity.

\begin{figure}[t]
	\includegraphics[width = 8.6cm]{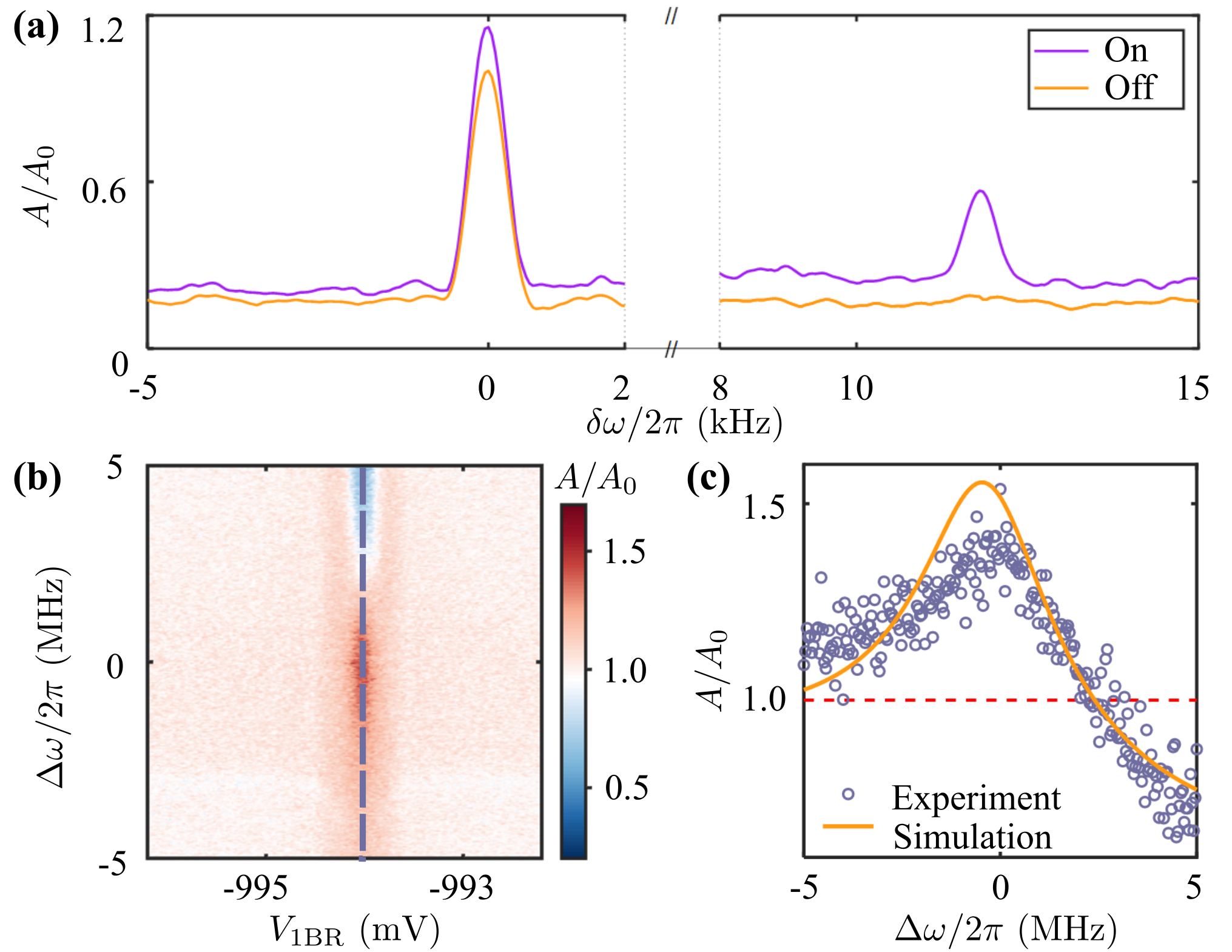} 
\caption{Characterization of the SAPA. (a) Spectra of the parametric amplification as a function of the probe frequency detuning with respect to the input signal $\delta \omega$. The orange line shows the signal without applying a pump tone, centered at $\omega_{s}=\omega_{r}$. The purple line represents the scenario with the pump tone applied to DQD$_1$, centered at $\omega_{s}=\omega_{r}-2\pi\times 4~\rm MHz$. (b) The normalized transmission amplitude under pump, as functions of frequency difference $\Delta \omega=\omega_p-\omega_s$ and gate voltage $V_{\mathrm{1BR}}$. (c) The signal with detuning $\varepsilon=0$, as marked by the gray dashed line in (b). The red dashed line indicates $A/A_0=1$.}
\label{Fig3}
\end{figure}

To further validate our physical model, the additional spectrum of transmitted signal is measured by a spectrum analyzer and plotted as a function of the frequency detuning with respect to the input signal ($\delta \omega$), as shown in Fig.~\ref{Fig3}(a). When the pump is off, we set the probe frequency at $\omega_{s}=\omega_{r}$ and tune DQD$_1$ to far detuning (orange line). We then tune DQD$_1$ to the proximity of zero detuning and activate the SAPA by turning on the pump tone. The probe frequency is set at $\omega_{s}=
\omega_{r}-2\pi\times 4~\rm MHz$, which corresponds to the maximum gain frequency as described above. Compared with the result without pump, the presence of an idler tone near $\delta \omega/2\pi\approx 12\,\mathrm{kHz}$ provides clear evidence of the non-degenerate four-wave mixing process of the DQD-based SAPA (see Supporting Information). The amplitude difference at $\delta \omega/2\pi\approx 0$ reveals a parametric amplification with an effective gain $G_{e}=1.99~\rm dB$. Due to the drift of the work point, the effective gain is smaller than that in Fig.~\ref{Fig2}. 

The tunability of the SAPA is characterized by measuring the amplitude $A/A_0$ as a function of the frequency difference $\Delta \omega$ between the pump and the signal tones. The signal frequency is fixed at $\omega_s-\omega_r=2\pi\times3\,\mathrm{MHz}$, while the pump frequency is scanned.Fig.~\ref{Fig3}(b) reveals that the gain is present only when the DQD is near resonance with the cavity mode. This observation confirms that the parametric amplification relies on the strong coupling between the DQD and the cavity, and that the gain is mediated by the hybridized qubit-cavity states. Notably, when $\varepsilon=0$ and the pump-signal detuning is smaller than the cavity resonance bandwidth, the effective gain exceeds unity, as shown in Fig.~\ref{Fig3}(c). This behavior can be attributed to the cavity mode enhancement of both the pump and signal fields, which is a unique feature of the DQD-based SAPA. As the frequency difference increases beyond the cavity bandwidth, the effective gain decreases and eventually becomes less than unity due to the pump and DQD induced cavity resonance shift. The SAPA gain center can be fine-tuned over $\sim 9~\rm MHz$ by adjusting the DQD parameters, sufficient for compensating cavity frequency variations. In addition, the 1\,dB gain compression point occurs at an input probe power of -120\,dBm (see Supporting Information).

Similarly, we also investigate the performance of DQD$_2$ as a parametric amplifier, and achieve a comparable parametric gain of 10.64\,dB and an effective gain of 3.72\,dB (see Supporting Information). The successful demonstration of parametric amplification using both DQDs highlights the reproducibility and robustness of our approach.

\section{On-chip readout and amplification}\label{section4}

The DQD-based SAPA featuring high gain, wide tunability, and compatibility with multiple DQDs opens up new opportunities for on-chip signal processing and enhanced readout. In our reconfigurable system, we first set DQD$_2$ as the target qubit to be readout while DQD$_1$ functions as the SAPA. We adjust the tunnel coupling of DQD$_2$ to the dispersive regime $2t_{c}^{(2)}/2\pi\hbar =\rm5.8~GHz$, and characterize the readout performance by measuring the transmittance contrast of on-resonance transmission amplitude. This continuous measurement not only allows for quantitative estimation of the SNR, but also can avoid the calibration of the qubit state excitation.

\begin{figure}[b] %t!
\includegraphics[width = 8.6cm]{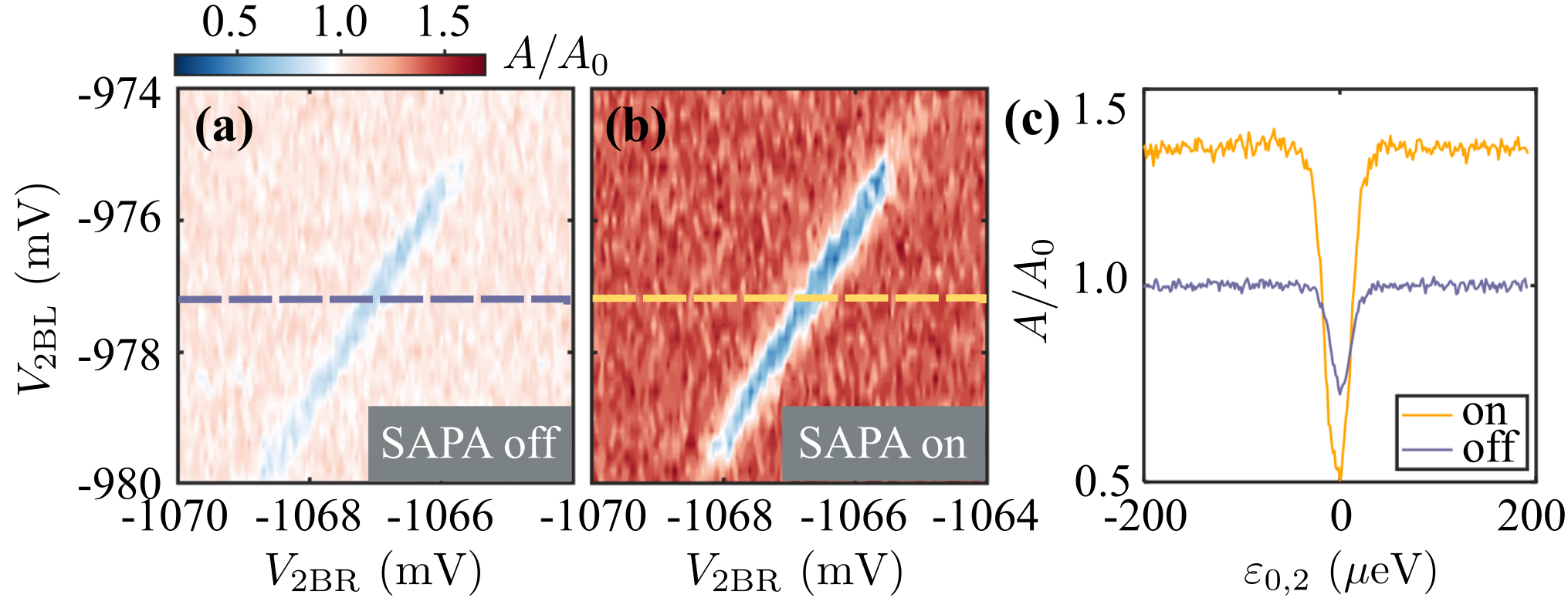}%[scale=0.35] 
%6.7GHz $\rm -978.9 ~ mV$
\caption{On-site readout enhancement using the SAPA. (a) and (b), Charge stability diagrams of DQD$_2$, without and with a pump tone applied to DQD$_1$, respectively. (c) Average transmission spectra of DQD$_2$ as a function of the detuning $\varepsilon_{0,2}$, with (orange) and without (gray) a pump tune. The values of $V_{\mathrm{2BL}}$ are fixed as indicated by dashed lines in (a) and (b).}
 \label{Fig4}
\end{figure}

Figure \ref{Fig4}(a) shows a charge stability diagram of DQD$_2$. Without SAPA by pump turn-off and fixed probe frequency at $\omega_s=\omega_r$, the tunneling line of DQD$_2$ is barely visible due to the low SNR. The gray line in Fig.~\ref{Fig4}(c) illustrates the lines cut along $V_{\rm 2BL} \rm =-977.1 ~ mV$. 

In contrast, when DQD$_1$ is tuned to the parametric amplification regime, and the pump tone is activated, the cavity transmission exhibits a pronounced dispersive shift, as shown in Fig.~\ref{Fig4}(b). Here, the probe frequency is set to $\omega_s=\omega_r-2\pi\times4~\rm MHz$ to achieve maximum gain. From the lines cut in Fig.~\ref{Fig4}(c), the readout signal is significantly enhanced by SAPA from two aspects (see Supporting Information): first, the amplitude of the transmittance is enhanced, as revealed by the effective gain $G_e$ in the above demonstrations. Second, the extinction ratio is enhanced because $\kappa$ is reduced due to the gain.

Additionally, the SAPA inevitably introduces extra noise which elevates the background fluctuation level, as observed in Fig.~\ref{Fig3}(a). Based on the noise chain model \cite{Gardiner_Quantum_2004}, the added noise from this SAPA is approximately 1.5 quanta, corresponding to an effective temperature $T_{\rm eff} \approx 0.3~\rm K$ (see Supporting Information). The improvements in the readout achieved by the SAPA can be directly quantified by the amplitude SNR:
\begin{equation}
	\mathrm{SNR}=\frac{\overline{A_1}-\overline{A_0}}{(\sqrt{\Delta A_1^2+\Delta A_0^2})}.
	%\tag{2}
\end{equation}
Here, $\overline{A_1}$ and $\overline{A_0}$ are the amplitude average values of 30 repeated measurements at $\varepsilon_{2}\gg g$ and $\varepsilon_{2}=0$, respectively, and $\Delta A_1$ and $\Delta A_0$ denote the corresponding standard deviations of transmitted signal. The SNR is evaluated by varying the detuning $\varepsilon_{2}$ of DQD$_2$ while DQD$_1$ operates as the SAPA with fixed parameters. As shown in Fig.~\ref{Fig4}(c), when the SAPA is turned off, the extinction of signal output $\delta A$ of DQD$_2$ is 0.3, and the SNR is 5.5. Remarkably, when DQD$_1$ is operated as SAPA, $\delta A$ increases to 1.0, and the SNR is enhanced to 10.9. Similar measurements are performed by exchanging the role of the two reconfigurable DQDs, showcasing the flexibility and adaptability of our SAPA (see Supporting Information). Table \ref{Table1} summarizes the key results for different configurations, showing SAPA by either DQD$_1$ or DQD$_2$ can boost the SNR by $3\,\mathrm{dB}$.

Our further theoretical simulations reveal that the gain is influenced by the DQD-cavity coupling strength, cavity decay rate, and decoherence rate. Crucially, the gain and signal-to-noise ratio (SNR) can be significantly enhanced by optimizing the DQD-cavity coupling strength and the cavity decay rate. For instance, increasing $g_c^{(1,2)}$ to $2\pi\times70~{\rm MHz}$ or reducing $\kappa$ to $2\pi\times10~{\rm MHz}$ yields an impressive 10\,dB enhancement in effective gain and SNR with the same pump power (see Supporting Information). This can be accomplished by enhancing the cavity impedance and the coupling strength through careful positioning or structuring of the quantum dot coupling electrode~\cite{Scarlino_In_2022}. Additionally, integrating on-chip LC filters can help reduce cavity loss~\cite{Mi_Strong_2017,Xu_On_2024}, further improving the SAPA performance. Beyond these device-level optimizations, exploring advanced spin-charge hybrid modes such as the ``RX qubit" implemented in a triple quantum dot, represents another promising direction for future improvements~\cite{Landig_Coherent_2018, Jiang_Coupling_2025}. This strategy aims to maintain a large electric dipole moment for strong coupling while reducing the sensitivity to charge noise.

\begin{table}[h!]
	
	\caption{\label{Table1} \textbf{Comparison of signal extinction $\delta A$ and SNR}}
	\centering
	\setlength{\tabcolsep}{8.5mm}
	\begin{tabular}{rrrrrrrr}
		
		\hline
		
		\multicolumn{2}{c}{SAPA, On/Off}&{$\delta A $}&{SNR}\\
		\hline 
		\multicolumn{2}{c}{DQD$_1$, On}&1.0&10.9 \\
		\multicolumn{2}{c}{DQD$_1$, Off}&0.3&5.5 \\
		\multicolumn{2}{c}{DQD$_2$, On}&0.7&8.0 \\
		\multicolumn{2}{c}{DQD$_2$, Off}&0.2&3.6 \\

		\hline
	\end{tabular}
\end{table}

\section{Discussion and conclusions}\label{section5}

The successful demonstration of on-chip readout enhancement using the DQD-based SAPA, paves the way for further improvements and extensions of this technique. While we have not yet performed measurements in a large magnetic field for spin qubit operation, the device is inherently compatible with such fields, due to the robustness of both the quantum dot charge state and the superconducting cavity~\cite{Samkharadze_High_2016}. We demonstrate a reconfigurable quantum circuit architecture where DQDs can serve either as qubits or amplifiers within the same platform, which suggests a hybrid architecture comprising a cavity and multi-qubits with on-site low-noise amplifiers, offering a potential platform for scalable quantum applications. To ensure its scalability, future research will investigate strategies such as frequency multiplexing or novel architectures that separate the coupling and readout functions, thereby mitigating the risk of frequency crowding. In addition, scaling to a multi-SAPA architecture is a promising direction. Potential crosstalk can be managed by combining the strategy of operating each SAPA at a distinct frequency with active calibration using a multi-tone drive, providing a clear pathway to scalable, high-fidelity quantum readout chains.

Beyond its technological implications, the reconfigurable DQD-based quantum circuit provides a powerful tool for exploring fundamental quantum physics and implementing novel quantum protocols. Analogous to superconducting systems, the SAPA promises the realization of single-shot measurements of semiconductor spin qubits, a crucial capability for quantum error correction and feedback control. The SAPA-enhanced dispersive readout also opens avenues for quantum precision measurement, enabling the detection of external electric fields through DQDs or the measurement of force and displacement through piezoelectric coupling to nanomechanical resonators~\cite{Ockeloen_Stabilized_2018}. Furthermore, the strong nonlinearity and high cooperativity of the DQD-cavity system could be harnessed to generate non-classical states of the cavity~\cite{Schuster_Resolving_2007}. By working at the degenerate regime, the SAPA can provide on-site squeezing interactions for enhanced qubit readout and qubit-qubit entanglement~\cite{Gu_Microwave_2017}.

\begin{acknowledgments}
This work was funded by the National Natural Science Foundation of China (Grant Nos. 92265113, 12574552,12474490, 12034018, and 92265210) and the Quantum Science and Technology-National Science and Technology Major Project (Grant No. 2021ZD0302300 and 2021ZD0300203). CLZ was also supported by the Fundamental Research Funds for the Central Universities, and USTC Research Funds of the Double First-Class Initiative. The numerical calculations in this paper have been done on the supercomputing system in the Supercomputing Center of University of Science and Technology of China. This work was partially carried out at the USTC Center for Micro and Nanoscale Research and Fabrication.
\end{acknowledgments}

\bibliographystyle{apsrev4-2}
\bibliography{Ref1} 

\end{document}